\documentstyle[prc,aps,epsf]{revtex}

\include{revtex}

\begin{document}

%\draft

%\preprint{}

\title{Influence of higher-order deformations in the \protect$^{34}$S 
+ $^{168}$Er fusion reaction}

\author{C. R. Morton, A. C. Berriman, R. D. Butt, M. Dasgupta, 
D. J. Hinde, A. Godley, and J. O. Newton}
\address{Department of Nuclear Physics, Research School of Physical 
Sciences and Engineering,\\ Australian National University, Canberra, 
ACT 0200, Australia}

\author{K. Hagino}
\address{Yukawa Institute for Theoretical Physics, Kyoto University, 
Kyoto 606-8502, Japan}

\date{Received 16 April 2001, Accepted 15 June 2001, Phys. Rev. C}

\maketitle

\begin{abstract}

The shape of the measured barrier distribution for the 
$^{34}$S + $^{168}$Er reaction is analysed using the coupled-channels 
description.
The $^{168}$Er nucleus is a good candidate to test current 
fusion models description of deformation since it has a large quadrupole 
deformation with an insignificant hexadecapole deformation.  
Coupling to weaker channels, the $2^+_1$ state in 
$^{34}$S, the $3^-_1$ state in $^{168}$Er, and the pair neutron transfer 
channel, all were found to have little influence on the barrier 
distribution.
A successful description of the barrier distribution was only obtained 
after the hexacontatetrapole deformation term in $^{168}$Er 
($\beta_6^T$) was included in 
the coupling scheme.  However, a positive value for $\beta_6^T$ was needed 
where the macroscopic-microscopic model predicts a negative one.  

\end{abstract}


\section{Introduction}
\label{intro}

The influence of the internal structure of nuclei
on the probability for heavy-ion fusion is well established.
Coupling of the relative motion of projectile and target 
nuclei to internal nuclear degrees of freedom produces a distribution of 
fusion barriers that replaces the single fusion barrier which results 
when these couplings are not 
included~\cite{Wong72,Vaz74,Dasso83,Rowley91}.  
The degrees of freedom that affect fusion include the 
deformation of one or both reactants, the excitation of collective 
surface vibrational modes, and single or multiple-nucleon transfer 
channels.

In general, collective modes involving large numbers of nucleons play
the major role in determining the shape of the fusion barrier 
distribution.  Evidence for this can be seen in the measured barrier 
distribution for the system $^{16}$O + 
$^{144}$Sm~\cite{Morton94,Leigh95}, for example, where the lowest 
energy octupole state in $^{144}$Sm accounted for a large part 
of the structure in the experimental barrier distribution, whilst the 
transfer channels had only a relatively minor effect at energies in the 
barrier region.  This difference is due to the weakness of the coupling 
of one- or two-particle transfer channels relative to the coupling 
strength of the collective modes~\cite{Broglia85}.
This does not imply, however, that there are no significant effects of
transfer on fusion.
The effects of positive (effective) $Q$-value transfers, when present, 
are most evident in the enhancement of the fusion cross section at 
energies well below the average fusion barrier $B_0$, since they 
determine the energy of the lowest 
barrier~\cite{Dasso83,Leigh95,Dasgupta98}. 

\begin{figure}
\epsfxsize=15cm
\epsfbox[30 90 510 765]{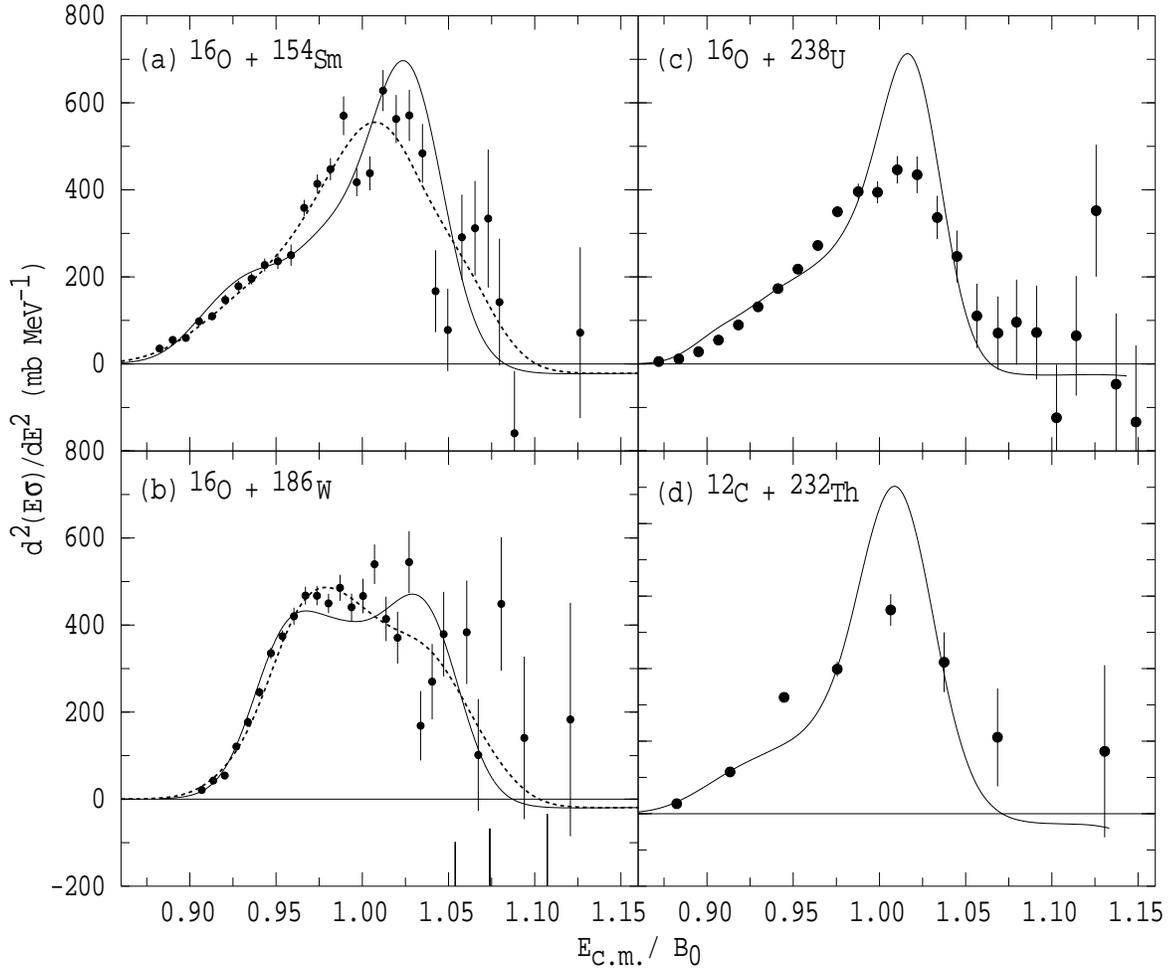}
\caption{Fusion barrier distributions for a range of deformed 
nuclei in the reactions (a) 
\protect$^{16}$O + $^{154}$Sm~\protect\cite{Leigh95,Wei91}, 
(b) \protect$^{16}$O + $^{186}$W~\protect\cite{Leigh95,Lemmon93}, 
(c) \protect$^{16}$O + $^{238}$U~\protect\cite{Hinde95}, and (d)
\protect$^{12}$C + $^{232}$Th~\protect\cite{Mein97}.   } 
\label{fig1}
\end{figure}

Two examples of the sensitivity of the fusion barrier distribution 
to the effects of deformation are the 
$^{16}$O + $^{154}$Sm~\cite{Leigh95,Wei91} and $^{16}$O + 
$^{186}$W reactions~\cite{Leigh95,Lemmon93}.
Calculations which could reproduce 
the measured barrier distributions required not only
a quadrupole deformation term but also inclusion of a hexadecapole term, 
where the sign of the latter was responsible for the difference in 
shapes of the two barrier distributions [compare the solid lines in 
Fig.\ \ref{fig1}(a) and (b)].
Although the quadrupole and hexadecapole degrees of freedom are 
responsible for a large part of the barrier distribution's shape, there 
are small differences between the shape of the theoretical and 
experimental barrier distributions at centre-of-mass energies 
$E_{\mbox{\scriptsize c.m.}} \lesssim B_0$.
It was shown in Ref.\ \onlinecite{Leigh95} however, that if additional
coupling to vibrational states and transfer channels were included in 
the theoretical model, then a better reproduction of the shape of the
measured barrier distributions resulted.  The calculations of 
Ref.\ \onlinecite{Leigh95} are shown by the dashed lines in 
Fig.\ \ref{fig1}(a) and (b) for the $^{154}$Sm and $^{186}$W targets, 
respectively.  The deformation parameters obtained from these
fits to the data were found to be in closer agreement with published 
(non-fusion) values~\cite{nonfusion} than those calculations which 
excluded the additional couplings.  

However, as recognised in Refs.\ \cite{Leigh95,Lemmon93}, the addition 
of {\em any} weak couplings may result in an improvement in 
the reproduction of the shape of the measured barrier distribution.  
The resulting deformation parameters from the fit to the fusion data 
are in better agreement with non-fusion values since, to a large extent, 
the additional couplings in $^{154}$Sm and $^{186}$W mimic the changes 
in the deformation parameters required to reproduce the barrier 
distribution shape. 

In other measurements involving deformed target nuclei, a discrepancy 
between the theoretical and experimental barrier distributions in the 
region $0.94 \lesssim E_{\mbox{\scriptsize c.m.}} / B_0 \lesssim 1.0$ 
MeV was found in possibly four different systems.
Examples are shown in Fig.\ \ref{fig1}(c) and (d) for the 
$^{16}$O + $^{238}$U~\cite{Hinde95} and $^{12}$C + 
$^{232}$Th~\cite{Mein97} reactions, respectively. 
Is this discrepancy due to coupling to weaker channels or to some 
deficiency in the theoretical description of fusion involving deformed 
nuclei?  Or are the measured fusion barrier distributions sensitive to 
higher-order deformations, such as the $\beta_6^T$ deformation, presently 
absent from the analyses of fusion involving deformed nuclei?
A recent paper~\cite{Rumin99} addressed this last question by examining 
the influence of the $\beta_6^T$ deformation on fusion for the 
$^{16}$O + $^{154}$Sm, $^{186}$W, and $^{238}$U systems.
However, the disadvantage of this analysis~\cite{Rumin99} was that all 
the nuclei involved in the study have large $\beta_4^T$ deformations, 
which meant it was difficult to isolate the effects due to the 
$\beta_6^T$ term from the $\beta_4^T$ contributions.

In order to investigate the influence of higher-order deformations, 
such as the $\beta_6^T$ term, a target nucleus with a pure prolate 
deformation should be chosen, to avoid distortion due to a 
hexadecapole component.  
The $^{168}$Er nucleus is a good candidate since it has 
a large prolate deformation and a hexadecapole deformation that is 
expected to be zero, or very small~\cite{Hendrie68,Nilsson69}.
The $\beta_3^T$ vibrational mode in $^{168}$Er is also weakly coupled
compared to the $^{238}$U target nucleus where this mode is significant.
Additionally, it is desirable to choose a reaction that involves a 
projectile nucleus with a high charge because the width of the barrier 
distribution is proportional to $Z_1 Z_2$, which produces a 
``magnification'' of the coupling effects.

The reaction $^{34}$S + $^{168}$Er is suitable based on the above 
criteria.  It was recently used~\cite{Morton00let,Morton00} in a 
measurement of the fusion cross sections, and the fusion barrier 
distribution, for input into transition state model calculations 
used to test fission models at large angular momenta.  
In this paper, the fusion barrier 
distribution for $^{34}$S + $^{168}$Er from 
Refs.\ \cite{Morton00let,Morton00} is analysed to examine the role 
higher-order deformations play in heavy-ion fusion.
Details of the experiment and results for the $^{34}$S + $^{168}$Er 
measurement have been published~\cite{Morton00}.  
In Sec.\ \ref{diffuse} that follows, the sensitivity of the 
$^{34}$S + $^{168}$Er barrier distribution to the nuclear potential 
parameters is examined.  Then the effects of deformation on 
the barrier distribution are calculated in Sec.\ \ref{deform} after 
checking the adequacy of the approximate treatment of the excitation 
energies of states in $^{168}$Er.
In Sec.\ \ref{other}, the effects of coupling to the 
octupole vibration in $^{168}$Er, and states in $^{34}$S is examined, 
followed by a discussion on the effects of transfer couplings, and 
the conclusion in Sec.\ \ref{conclusion}.

\section{NUCLEAR POTENTIAL PARAMETERS}
\label{diffuse}

One uncertainty concerning the theoretical description of heavy-ion
fusion is the true form of the nuclear potential between interacting
nuclei in the absence of channel coupling.  In fusion analyses, an 
energy-independent nuclear potential of Woods-Saxon form has often 
been used, with
\begin{equation}
 V(r)=-\frac{V_0}{1+\exp[(r-r_0 A_P^{1/3}-r_0 A_T^{1/3} )/a] }, 
\label{eq1}
\end{equation}
where $V_0$ is the depth, $r_0$ is the radius parameter, and $a$ is the 
diffuseness of the nuclear potential.  One method for determining the 
parameters of the Woods-Saxon potential is the procedure described in 
Ref.\ \onlinecite{Leigh95}, where the fusion cross sections $\sigma$ 
have been fitted with a single barrier penetration model at energies 
$\approx 10\%$ above the average barrier where there is no 
longer any significant barrier strength.
The rationale behind this procedure is that at energies well above 
the average barrier the calculated cross sections are very 
sensitive to $B_0$ but relatively insensitive to the channel couplings, 
thus enabling an estimate of the ``bare'' nuclear potential.  

With the advent of precise fusion data, even this small sensitivity 
to the channel couplings at higher energies was measurable, and was 
dealt with usually in one of two ways.
In calculations that included couplings with excitation energies 
smaller than the curvature of a single barrier $\hbar \omega_0$, 
a small increase in the fusion barrier was made in order to retain 
the quality of the fit to the high energy data obtained without the 
couplings.  Where the excitation energies of states was greater 
than $\hbar \omega_0$, they were not included explicitly in the 
calculations, since these states only contribute to a renormalisation 
of the ``bare'' nuclear potential and have no effect on the shape of 
the barrier distribution~\cite{Hagino97}.

If these higher lying states are included in the calculations, care 
should be taken not be count their effects twice, or wrong conclusions 
about the position of average fusion barrier will be drawn.  An example
of this problem is shown in Ref.\ \cite{Stefanini00} where the quoted
``average barrier'' is several MeV above the {\em measured} average 
fusion barrier as determined by their experimental fusion barrier 
distribution.

\begin{figure}
\epsfxsize=19cm
\epsfbox[80 50 510 465]{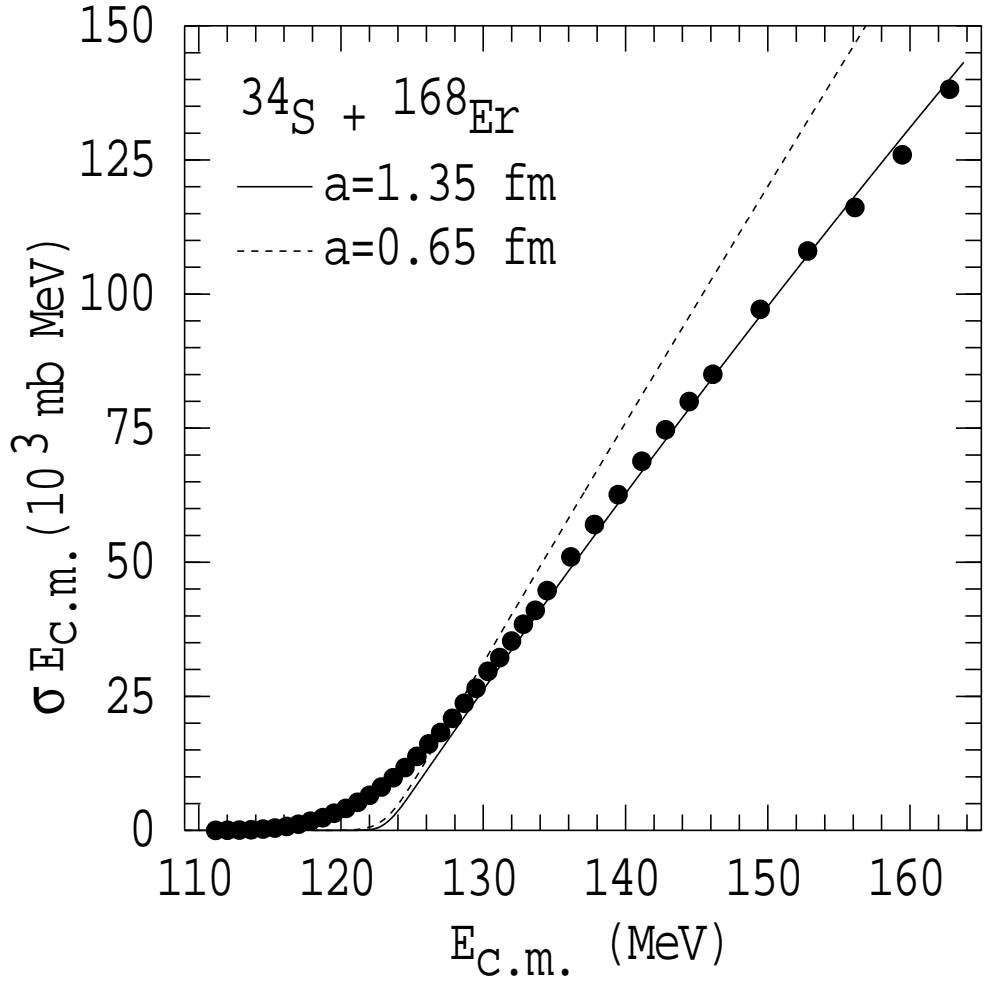}
\caption{Linear plot of 
\protect$\sigma E_{\mbox{\scriptsize c.m.}}$ for the measured 
fusion cross-sections (solid points) for 
\protect$^{34}$S + $^{168}$Er compared with two single-barrier 
penetration model calculations, one with \protect$a=1.35$
fm (solid line), and the other with \protect$a=0.65$ fm 
(dashed line). }
\label{fig2}
\end{figure}

The fit to the high-energy fusion cross-sections 
for $^{34}$S + $^{168}$Er is shown by the
solid line in Fig.\ \ref{fig2}, where the quantity 
$\sigma E_{\mbox{\scriptsize c.m.}}$ has been plotted on a linear
scale for clarity.
The parameters for the nuclear potential obtained from this fit 
are given in the second row of Table~\ref{table1}, noting that the 
diffuseness obtained was $a=1.35$ fm.  The resulting average barrier 
is $B_0=123.1$ MeV at a barrier radius of $R_B=11.1$ fm.
The diffuseness from this fit is much larger than the results from 
fits to elastic scattering data in Ref.\ \onlinecite{nonfusion}, where 
the fitted value for the diffuseness was found to be $a=0.63$ fm.  
Large values for $a$, in the range $a=0.84$--$1.35$ fm, 
have also been required to fit data in recent 
measurements~\cite{Leigh95,Morton99,Newton00} involving a range of 
projectile-target combinations.  In Ref.\ \onlinecite{Newton00} 
an alternative form for the nuclear potential was suggested as a 
possible explanation for the large values of $a$.  If the nuclear 
potential fell more rapidly with increasing values of $r$ than 
the rate suggested by the Woods-Saxon form of Eq.\ \ref{eq1}, 
it may be possible to match the potential at large $r$ required
to fit the elastic scattering data, whilst retaining the fit to 
fusion cross-sections at the smaller values of $r$ probed by fusion 
collisions.

A clue as to the actual value for the diffuseness of the nuclear 
potential can be obtained from the slope of the fusion excitation
function in the tunnelling regime.  That is, at energies below the 
lowest barrier, where coupling effects no longer influence fusion.  
If the diffuseness obtained from fits to elastic scattering is
appropriate for fusion, then the fusion excitation function will also
have this slope provided the energy is sufficiently below the lowest
barrier.  If the fusion excitation function falls more 
rapidly with energy than a calculation with 
$a=0.63$ fm, this means fusion takes place through a wider barrier
(less penetrability), implying a larger value for $a$.
In the present case, it is difficult to determine the slope of the 
fusion excitation function because of the large width of the 
$^{34}$S + $^{168}$Er barrier distribution, resulting from the 
target nucleus deformation, and the possible influence of positive 
$Q$ value transfers.  However, other experiments have recently being 
performed~\cite{anu}, using reactions that have a suitably narrow 
barrier distribution, and preliminary analysis of this data does 
support a value for $a$ significantly larger than the elastic 
scattering results.

\begin{figure}
\epsfxsize=18cm
\epsfbox[80 50 510 465]{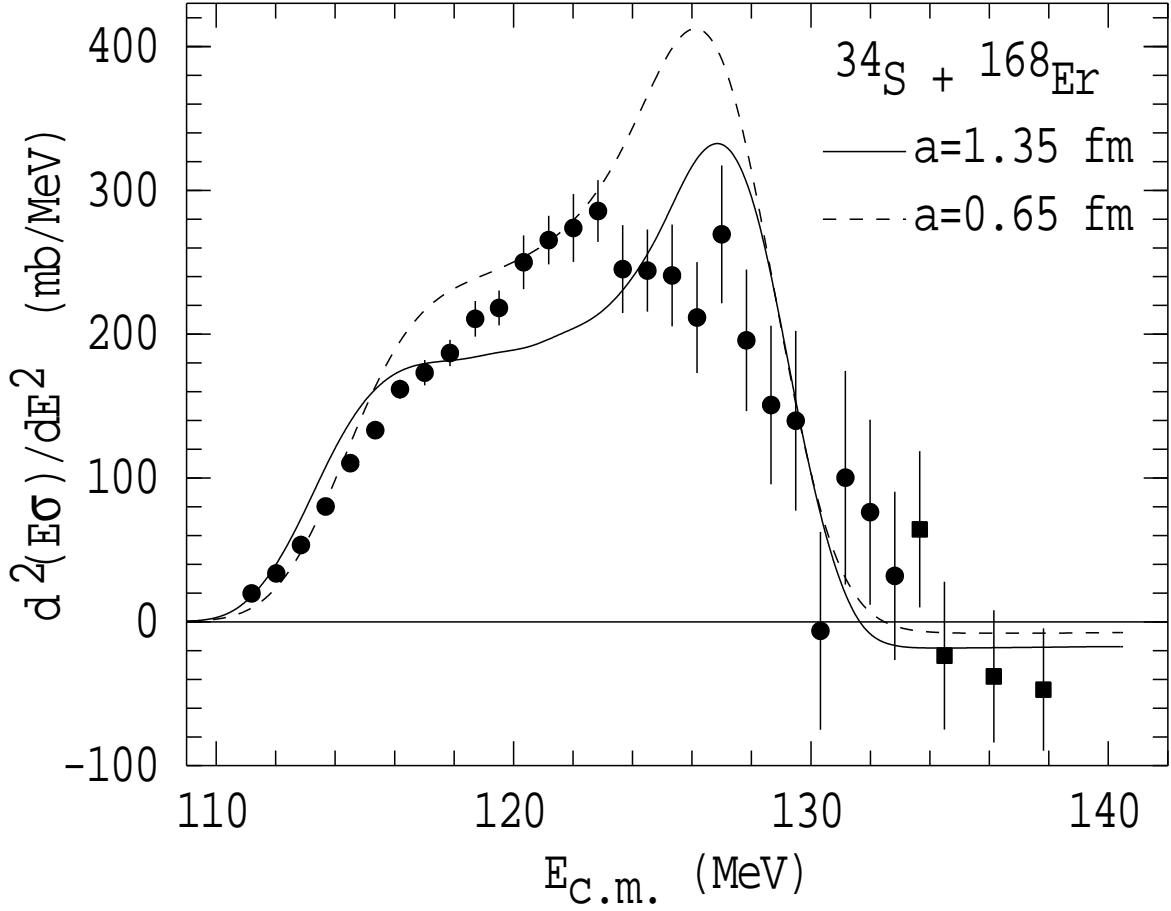}
\caption{The measured fusion barrier distribution for 
\protect$^{34}$S + $^{168}$Er~\protect\cite{Morton00}.  Here the 
second derivative was evaluated with a step length \protect$\Delta 
E_{\mbox{\scriptsize c.m.}} = 3.33$ MeV (solid circles) or 
\protect$\Delta E_{\mbox{\scriptsize c.m.}}= 6.66$ MeV (solid 
squares).  Also shown are two CC calculations which are identical 
except for the nuclear potential parameters, the solid line results 
from the calculation made with \protect$a=1.35$ fm, and the dashed 
line with \protect$a=0.65$ fm. See text. }
\label{fig3}
\end{figure}

A calculation was performed to demonstrate the effect of using a 
diffuseness smaller than the fusion data require.  A value of $a=0.65$
fm was chosen, which overestimates the measured cross-sections at high
energies (see the dashed line in Fig.\ \ref{fig2}).  
The nuclear potential parameters for this calculation are given in the 
third row of Table~\ref{table1}.  The potential parameters were chosen 
to keep the fusion barrier unchanged at $B_0=123.1$ MeV, which meant 
that $R_B$ was increased from $11.1$ to $12.0$ fm to compensate.  The 
fusion barrier distribution resulting from a calculation with $a=0.65$ 
fm and a quadrupole deformation of $\beta_2^T=0.338$~\cite{Raman87} 
only is shown by the dashed line in Fig.\ \ref{fig3}.  The barrier 
distribution from this 
calculation also leads to a deterioration in the agreement with 
measured barrier distribution.

\section{EFFECTS OF DEFORMATION ONLY}
\label{deform}

Having established that a smaller diffuseness parameter
made the agreement between the calculated and experimental 
excitation function and barrier distribution worse, 
the effects of deformation are examined in 
this Section.  Coupled-channels (CC) calculations were performed
with the code {\small CCDEGEN}~\cite{Haginounpub}, which is based on a 
version of the code {\small CCFULL} described in 
Ref.\ \onlinecite{Hagino99}.  In {\small CCDEGEN}, the effects of 
deformation are calculated by coupling to the ground-state 
rotational band of the deformed target nucleus, using the no-Coriolis 
approximation~\cite{Lindsay84,Tanimura85} with the excitation 
energies of the states in the ground-state band taken to be zero.
The no-Coriolis approximation has been shown to be adequate for reactions 
involving heavy-ions~\cite{Morton99,Tanimura87}.
These approximations enable the CC equations 
to be decoupled, and the resulting eigenchannel equations, which 
correspond to fusion of the (inert) projectile and deformed target 
nucleus whose symmetry axis is orientated at an angle $\theta$ with 
respect to the beam axis, are solved~\cite{Rumin99} to obtain the 
tunelling probability $P_J(E_{\mbox{\scriptsize c.m.}},\theta)$ for 
each partial wave $J \hbar$.  The fusion cross section is then 
calculated using
\begin{equation}
\sigma (E_{\mbox{\scriptsize c.m.}}) = \pi \lambdabar^2 \sum_{J} 
(2J+1) P_J (E_{\mbox{\scriptsize c.m.}}), 
\label{xs}
\end{equation}
where $P_J (E_{\mbox{\scriptsize c.m.}})$ is the total tunnelling 
probability for each $J$ averaged over all orientations, and is given 
by
\begin{equation}
P_J (E_{\mbox{\scriptsize c.m.}})=\frac{1}{2} \int_{0}^{\pi} 
P_J(E_{\mbox{\scriptsize c.m.}},\theta) \sin \theta d\theta.
\label{p}
\end{equation}
Equation~\ref{p} is exact for the classical situation, where the 
number of (degenerate) states in the rotor tends to 
infinity~\cite{Nag86}.  In actual rotational nuclei, the number of 
states is finite, and the 
integral in Eq.\ \ref{p} is evaluated up to some maximum value of spin 
$I_{\mbox{\scriptsize max}}$ (where all spins are even)~\cite{Nag86}.
This is done using an $n$--point Gaussian integration formula, where
$n/2=(I_{\mbox{\scriptsize max}}+2)$.  
For the calculations described in this work, for systems other 
than $^{34}$S + $^{168}$Er [see Fig.\ \ref{fig1}(c) and (d)], a total of 
$5$ states up to the $8^+$ state were included, which corresponds to a 
coupled-channels calculation with 
$5$ channels.  Including states with spins higher than the $8^+$ state 
did not make an appreciable difference to the calculated fusion 
cross section.  For the $^{34}$S + $^{168}$Er 
calculations, $6$ states, up to the $10^+$ state, were needed to 
evaluate the integral to the desired accuracy. 

\begin{figure}
\epsfxsize=14cm
\epsfbox[70 85 510 720]{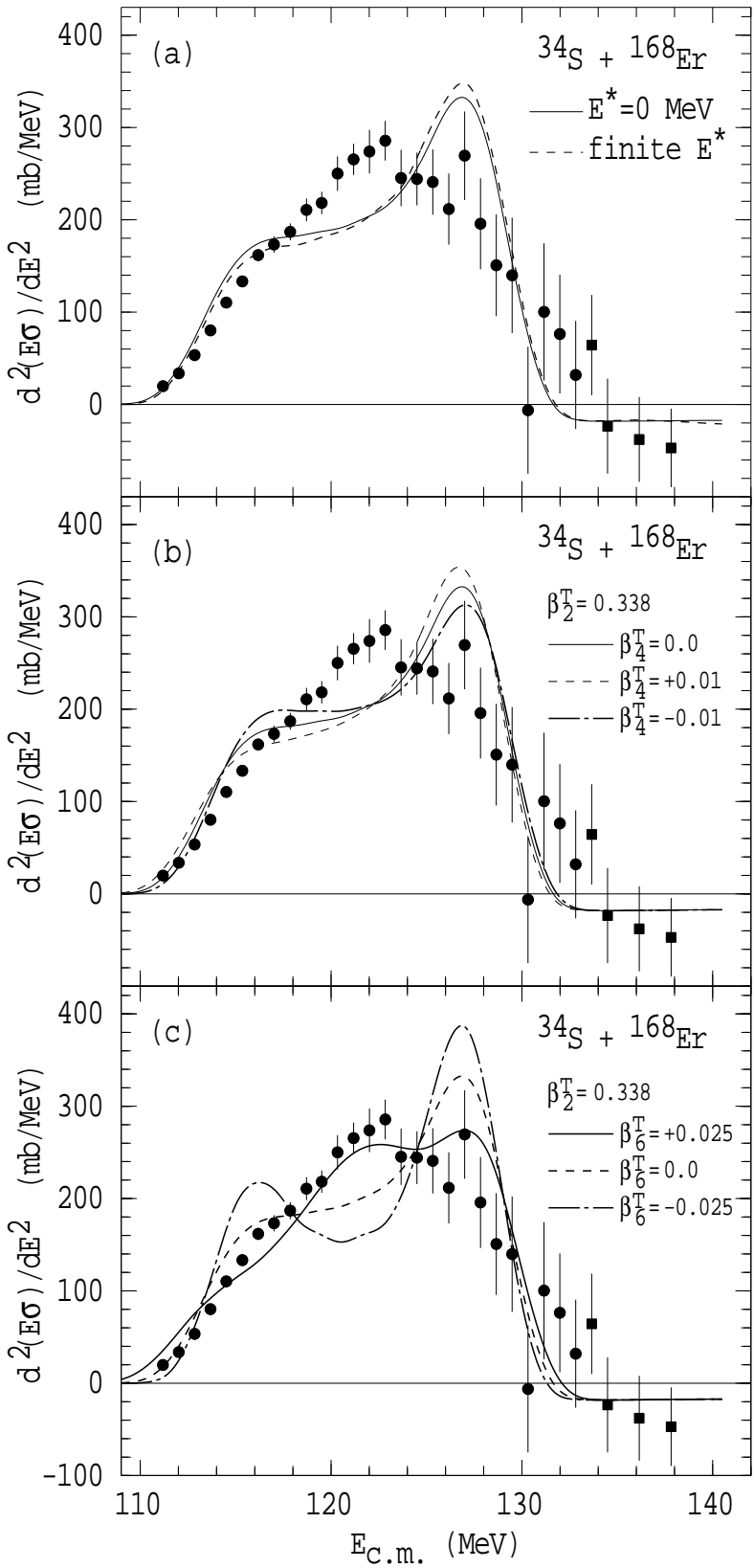}
\caption[]{(a) Test of the zero excitation energy approximation used in 
the CC code \protect{\small CCDEGEN} (solid line) compared with an exact 
calculation (dashed line).  See text for details. 
(b) Barrier distributions showing the effects of including the 
\protect$\beta_4^T$ deformation in addition to \protect$\beta_2^T$.  
The solid line is a \protect{\small CCDEGEN} calculation with 
\protect$\beta_4^T=0$, the dashed line with \protect$\beta_4^T=+0.01$, 
and the dot-dashed with \protect$\beta_4^T=-0.01$.  (c) Effects of the 
hexacontatetrapole deformation on the barrier distribution.  
Shown are \protect{\small CCDEGEN} calculations 
with \protect$\beta_2^T=0.338$ and \protect$\beta_6^T=0$ (dashed line), 
\protect$\beta_2^T=0.338$ and \protect$\beta_6^T=+0.025$ (solid line), 
and \protect$\beta_2^T=0.338$ and \protect$\beta_6^T=-0.025$ 
(dot-dashed line).  Note that \protect$\beta_4^T$ is zero in each of 
the calculations shown in panel (c).  }
\label{fig4}
\end{figure}

To test the zero excitation energy approximation used in these 
calculations, a comparison was made with a calculation which takes 
into account the finite excitation energies of the rotational 
states.  The calculation shown by the dashed line in 
Fig.\ \ref{fig4}(a) was made with the CC code 
{\small CCFULL}~\cite{Hagino99}, which included
coupling to all orders in the deformation parameter for the nuclear
coupling potential, with the energy of the first $2^+$ in $^{168}$Er 
at $79.8$ keV.  The excitation energies of higher members of the 
rotational band were calculated according to the usual formula for a 
rotating rigid body.  The solid line in Fig.\ \ref{fig4}(a) is the barrier 
distribution from the {\small CCDEGEN} calculation, using the geometric
description of Eq.\ \ref{p}.  The parameters for the nuclear 
potential were identical in these two calculations (see the second 
row of Table~\ref{table1}).  Both calculations were made with 
$\beta_2^T=0.338$~\cite{Raman87} and included coupling to all orders in 
the deformation parameter for the nuclear coupling potential.  
The difference between the two calculations is small
which can be attributed to the large deformation, or equivalently the
low energy of the first excited state in $^{168}$Er, although the 
agreement is not perfect~\cite{Rumin00}.

Having shown numerically that the zero excitation energy approximation 
in {\small CCDEGEN} is a good approximation to better than the accuracy
of the data, it was used in all the fusion calculations that follow to
test the effects of various deformation parameters.

\subsubsection{Quadrupole deformation}

The shape of the barrier distribution that results from consideration
of the quadrupole deformation only (solid line in Fig.\ \ref{fig4}(a) 
with $\beta_2^T=0.338$) cannot account fully for the shape of the measured 
barrier distribution.  Although the area under the calculation (when 
$a=1.35$ fm) matches to within 
$2$--$4\%$\footnote{The range depends upon the limits taken
for evaluation of the integral.} the area 
under the measured barrier distribution, the latter peaks at a lower 
energy than the theoretical curve and has a more ``triangular'' shape.
Varying the magnitude of $\beta_2^T$ does not improve the agreement.

\subsubsection{Hexadecapole deformation}

Although the hexadecapole deformation for $^{168}$Er is expected to be 
very small, a calculation was performed to ascertain the influence of 
this degree-of-freedom on the shape of the barrier distribution.
The results of two CC calculations are shown by the broken lines in 
Fig.\ \ref{fig4}(b), one with $\beta_2^T=0.338$ and 
$\beta_4^T=+0.01$ (dashed line) and the other with 
$\beta_2^T=0.338$ and $\beta_4^T=-0.01$ (dot-dashed
line).  These two values for $\beta_4^T$ span the range of 
likely values for $^{168}$Er, as determined in 
Ref.\ \onlinecite{Moeller95} where a value of $\beta_4^T=-0.007$ was
estimated.  Of the 
resulting range of shapes shown in Fig.\ \ref{fig4}(b), none lead to a 
significant improvement in the agreement with the data. 

\subsubsection{Hexacontatetrapole deformation}

Since the effect of the hexadecapole deformation for the 
$^{168}$Er nucleus is small, what then is the effect of 
the hexacontatetrapole ($\beta_6^T$) degree of freedom on fusion?
Guidance as to the magnitude and sign of the $\beta_6^T$ deformation 
comes from theory.
The ground-state shapes for a large number of nuclei have been 
calculated by M\"{o}ller, Nix, Myers, and Swiatecki using a global
macroscopic-microscopic model~\cite{Moeller95}.  
They calculated a value of $\beta_6^T=-0.025$ for 
$^{168}$Er~\cite{Moeller95}, which is a factor $3.5$ times larger 
than the $\beta_4^T$ estimate for the same nucleus.
A {\small CCDEGEN} calculation with $\beta_6^T=-0.025$, in addition to 
the quadrupole deformation ($\beta_2^T=0.338$), is shown by the dot-dashed 
line in Fig.\ \ref{fig4}(c).  The $\beta_6^T$ deformation does have a 
significant effect on the shape of the barrier distribution, 
leading to more peaked shapes at each end of the barrier distribution.  
However, the inclusion of the negative hexacontatetrapole term worsens 
the agreement with the data.

Another calculation was performed but this time with the opposite
sign for the $\beta_6^T$ deformation.  This calculation, with
$\beta_2^T=0.338$ and $\beta_6^T=+0.025$, is represented by the solid line 
in Fig.\ \ref{fig4}(c).  The inclusion of the 
$\beta_6^T$ term with a positive instead of negative sign now improves 
the agreement with the measured barrier distribution, although all 
its features are still not completely reproduced.

In the recent work of Ref.\ \cite{Rumin99}, Rumin {\em et al.} 
investigated the influence of the $\beta_6^T$ deformation on fusion 
for the $^{16}$O induced reactions
on $^{154}$Sm, $^{186}$W, and $^{238}$U.  In that work the authors 
concluded that the {\em fits} to the measured barrier 
distributions were improved by inclusion of the $\beta_6^T$ terms, but 
in doing so obtained unphysically small $\beta_4^T$ values for 
these actinides, contrary to their known hexadecapole deformations.
It would be reasonable to conclude that the presence of the 
$\beta_6^T$ 
term compensates for the large reduction in the $\beta_4^T$ values.
A contrasting approach is taken in this work, where the ``known'' 
values of $\beta_2^T$ and $\beta_4^T$ for $^{168}$Er are used rather 
than allowed to vary as free parameters in a fit to the barrier 
distribution.  The basis for this approach is that the $\beta_2^T$ 
value has been determined experimentally~\cite{Raman87}, and the 
theoretical basis for $\beta_4^T$ being close to zero is well 
established, since Er is midpoint the region of $Z$ where $\beta_4^T$
changes from a positive value (Sm and Gd, for example) to a negative 
one (Yb and Hf)~\cite{Hendrie68,Nilsson69}.

An advantage the present study has over the analysis of Rumin 
{\em et al.}\ \cite{Rumin99} is that the $\beta_4^T$ for $^{168}$Er 
nucleus is approximately seven times smaller than the $\beta_4^T$ for 
the $^{154}$Sm and $^{238}$U nuclei (see Table~\ref{Table2}), implying
there is very little influence of the hexadecapole deformation on the
shape of the barrier distribution for $^{34}$S + $^{168}$Er.
In addition, the estimated~\cite{Moeller95} magnitude for the 
$\beta_6^T$ 
value for $^{168}$Er is around $1.5$ times the estimate for $^{238}$U 
and $5$ times the $\beta_6^T$ for $^{154}$Sm, increasing the possibility 
that fusion barrier distribution might be sensitive to the effects of 
this deformation parameter.  
When the deformation parameters for each nucleus are placed into the 
multipole expansion of the nuclear part of the coupling interaction, the 
hexacontatetrapole term for $^{168}$Er is $3.6$ times larger 
than its hexadecapole term, compared to $^{238}$U case where the 
hexacontatetrapole term is $3.4$ times {\em smaller} than its 
hexadecapole term.  So if the $^{16}$O + $^{238}$U barrier distribution
is sensitive to the presence of the $\beta_4^T$ in $^{238}$U, then it may
not be unexpected that the barrier distribution for $^{34}$S + 
$^{168}$Er is sensitive to the $\beta_6^T$ deformation in $^{168}$Er.

Although using the magnitude for $\beta_6^T$ from M\"{o}ller's estimate
produced a good description of the measured barrier distribution, the
sign had to be inverted in order to achieve this agreement.
In Ref.\ \onlinecite{Rumin99}, a similar problem arose, where the 
sign for $\beta_6^T$ was also found to be opposite to theoretical
predictions, after {\em fits} to the 
experimental barrier distributions for $^{16}$O + $^{238}$U and
$^{16}$O + $^{154}$Sm were performed.  An attempt was made to address 
this sign problem by re-fitting the barrier distributions with, in 
addition to the $\beta_2^T$, $\beta_4^T$, and $\beta_6^T$ terms, 
coupling to the octupole vibration of $^{238}$U and $^{154}$Sm.  
This calculation then matched the theoretically predicted sign for 
$\beta_6^T$ but, as discussed above, unphysically
small values for $\beta_4^T$~\cite{Rumin99} were obtained.  
In the case of $^{34}$S + $^{168}$Er, it was not possible to explain 
its sign problem by coupling to the $3^-_1$ state in $^{168}$Er, 
because this state couples so weakly (see Sec.\ \ref{other} below).

To check that the positive sign required is not due to truncation of 
higher-order terms in the Coulomb coupling equation, another calculation 
was performed which included the $(\beta_2^T)^3$ terms of the Coulomb 
interaction.  It was found that these terms had a very minor effect 
on the calculated barrier distribution for $^{34}$S + $^{168}$Er, 
corresponding to a change by a ``line width'' in the shape of the 
barrier distribution, and could not explain why a positive sign for 
$\beta_6^T$ was needed.

Whether or not the measured barrier distribution for $^{34}$S + 
$^{168}$Er can determine the sign of the $\beta_6$ deformation in
$^{168}$Er remains an open question.
The macroscopic-microscopic calculations of M\"{o}ller {\em et al.}
\cite{Moeller95} predict strong negative $\beta_6$ deformation in the
$N \approx 100, Z \approx 60$ region, however it is worth noting the 
comment in Ref.\ \onlinecite{Moeller95} that the behaviour of 
hexacontatetrapole deformation across the chart of the nuclides is 
less regular than that of the lower order, even 
($\beta_2$, $\beta_4$) multipole distortions.

Since the size of the deformation parameters extracted from the 
measured barrier distributions can depend upon the presence or absence 
of weaker couplings~\cite{Leigh95}, the effects of these channels are
investigated in the next Section.

\section{COUPLING TO WEAKER CHANNELS}
\label{other}

\begin{figure}
\epsfxsize=18cm
\epsfbox[70 90 510 600]{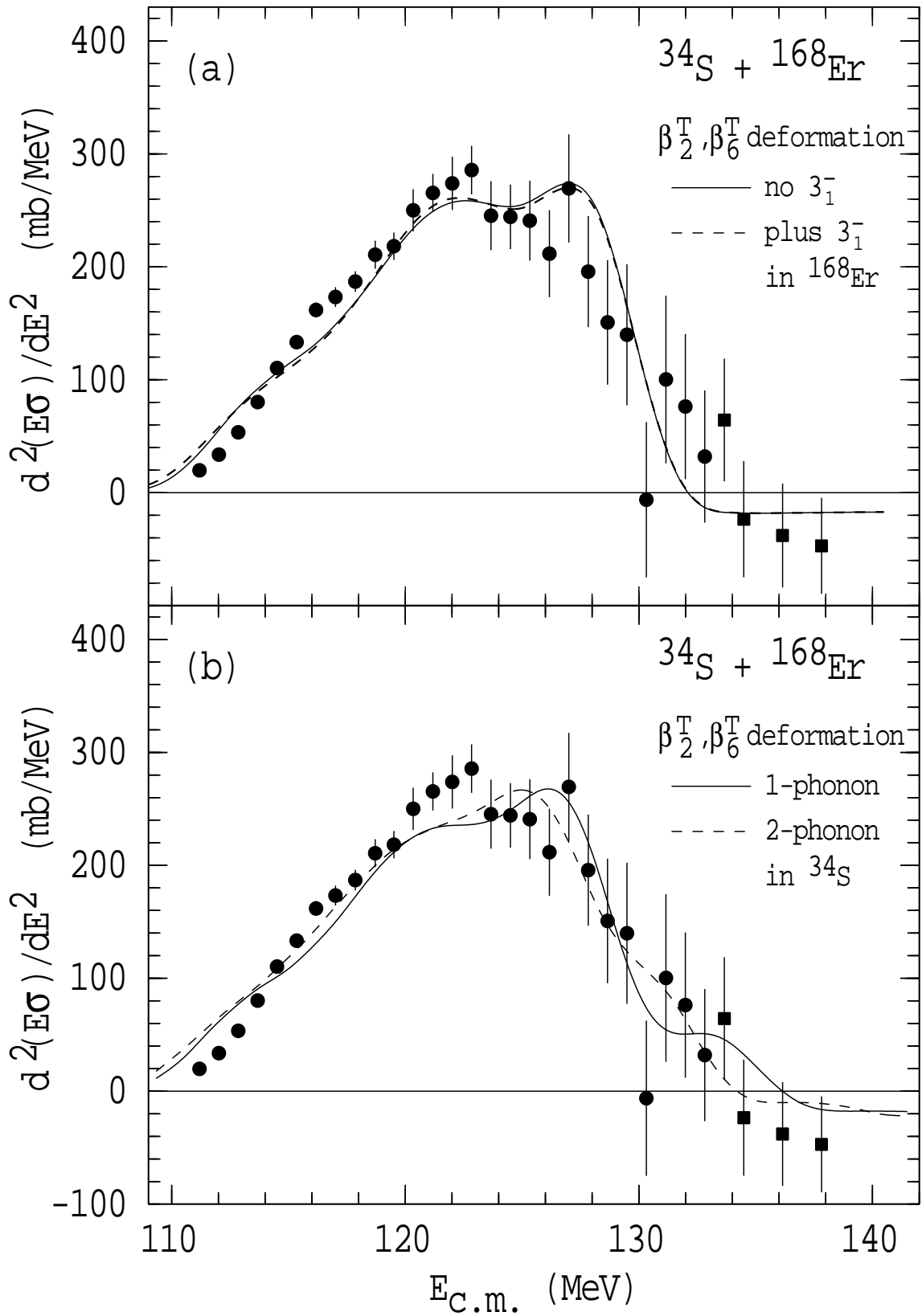}
\caption{(a) The effect on the barrier distribution when the 
\protect$3^-_1$ state in \protect$^{168}$Er is taken into account 
(dashed line), with \protect$\beta_2^T=0.338$ and 
\protect$\beta_6^T=+0.025$.  Also shown is the same 
calculation without the \protect$3^-_1$ state (solid line). 
(b) Barrier distributions for two calculations with
\protect$\beta_2^T$ and \protect$\beta_6^T$ deformation plus coupling 
to the \protect$2^+_1$ state in \protect$^{34}$S (solid line) and, 
in addition, coupling to the \protect$2$-phonon state in 
\protect$^{34}$S (dashed line).   }
\label{fig5}
\end{figure}

In this Section, coupling to channels other than the rotational coupling
is considered, starting with the octupole vibration in $^{168}$Er.  
The dashed line in Fig.\ \ref{fig5}(a)
is a {\small CCDEGEN} calculation with $\beta_2^T=0.338$, 
$\beta_6^T=+0.025$, and with the first $3^-$ state in $^{168}$Er.  
The energy for this state is $1431$ keV and the coupling strength was 
taken to be $\beta_3^T=0.064$, obtained from the measured 
$B(E3)\,\uparrow$ value~\cite{Spear89} with
a radius for $^{168}$Er taken as $1.06A_T^{1/3}$ fm~\cite{Leigh95}.  
The barrier distribution with the octupole coupling is only slightly 
different to the calculation without it [solid line in 
Fig.\ \ref{fig5}(a)] implying that this state does not play a
major role in the coupling scheme.  This is expected since the coupling 
strength for the octupole state is very weak.  Similarly, coupling to 
higher lying states, for example $\gamma$- and $\beta$-vibrations, should 
have very little influence on the barrier distribution, as shown to be 
the case in Ref.\ \onlinecite{Rumin99} for $^{16}$O induced reactions on 
the deformed $^{154}$Sm, $^{186}$W, and $^{238}$U targets.  
The effects of coupling to more exotic modes of 
excitation~\cite{Borner91}, such double-$\beta$ vibrations 
($2$-phonon collective excitations), have not been examined.  

\subsubsection{Coupling to states in \protect\mbox{$^{34}$S}}
\label{proj}

Up to this point in the analysis, $^{34}$S has been treated as 
inert in the calculations.  However, low lying states in $^{34}$S will 
contribute to the channel coupling, and the size of their influence 
should be calculated.  The barrier distribution shown by the solid line
in Fig.\ \ref{fig5}(b) is a calculation which includes $\beta_2^T$ and 
$\beta_6^T$ deformation in the target, plus coupling to the $2^+_1$ state 
in $^{34}$S, whose energy is $2.127$ MeV with a coupling strength of 
$\beta_2^P=0.25$.  
Also shown by the dashed line in Fig.\ \ref{fig5}(b) is the effect of 
coupling to the $2$-phonon excitation in the projectile 
(only the $[2^+_1 \otimes 2^+_1]$ $2$-phonon state was included). 
Both these calculations have been shifted up in energy by 
$1.0$ MeV to account for the adiabaticity of the projectile 
coupling~\cite{Morton99,Hagino97b}.  Coupling to the first excited state 
in the projectile does not dramatically alter the shape of the barrier 
distribution from the calculation which includes target excitations only.
The single-phonon excitation in the target produces a second peak in 
the barrier distribution centered around $133$ MeV, but the uncertainty
on $d^2(E_{\mbox{\scriptsize c.m.}}\sigma)/d 
E^2_{\mbox{\scriptsize c.m.}}$ is too large to identify such a feature.

\begin{figure}
\epsfxsize=18cm
\epsfbox[80 50 510 465]{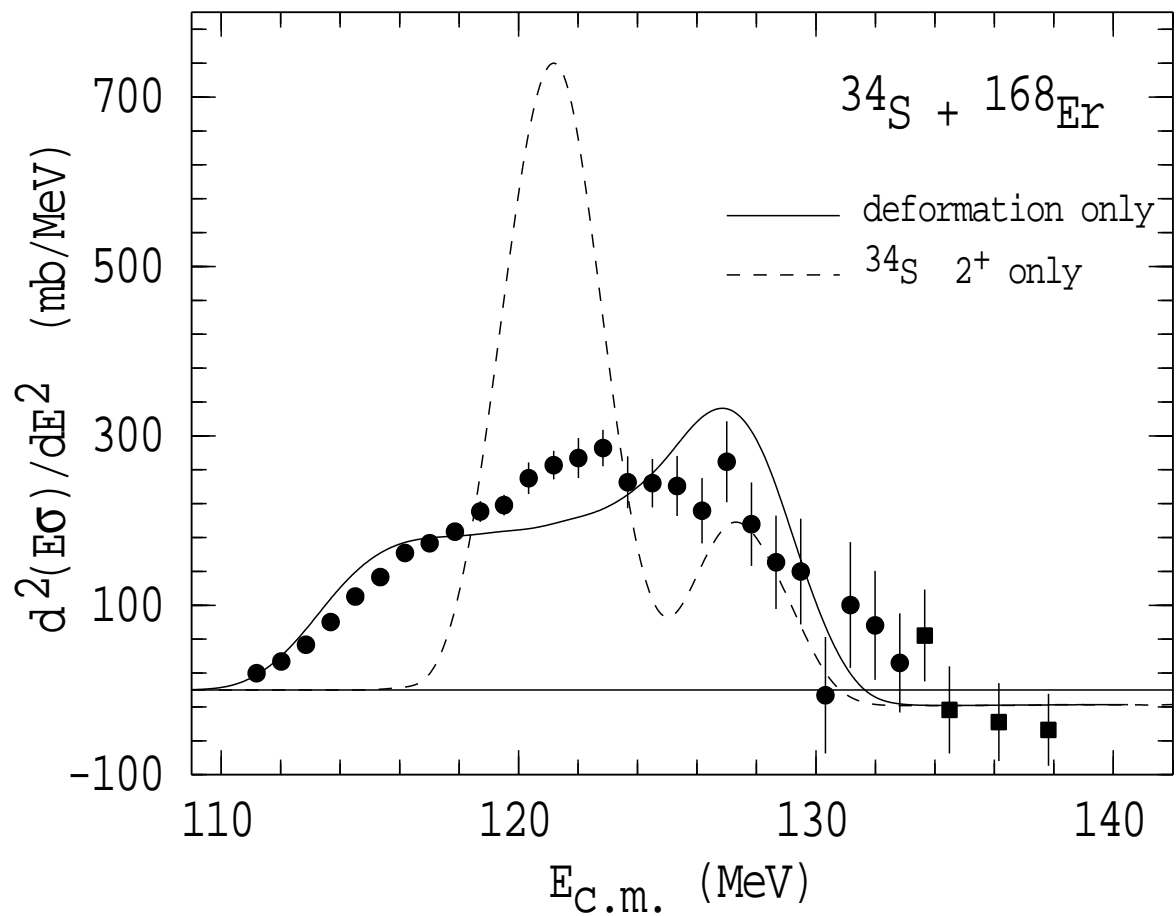}
\caption{Comparison of the barrier distribution that results from 
coupling to the first \protect$2^+$ state in \protect$^{34}$S only 
(dashed line) and target deformation only (solid line).  }
\label{fig6}
\end{figure}

Coupling to the first excited state in $^{34}$S has a small effect
because the strong coupling from $^{168}$Er produces a very 
broad barrier distribution.  This means that the overall change in 
the shape of the barrier distribution is small, even though projectile 
coupling by itself produces a significant feature.  This can be seen 
in Fig.\ \ref{fig6}, where a calculation with projectile coupling 
only is compared to a calculation which considers the target 
deformation only.

Two other calculations were performed, again with $\beta_2^T=0.338$ 
and $\beta_6^T=+0.025$ for $^{168}$Er.  The first included coupling to 
the $3^-_1$ state in $^{34}$S at an excitation energy of $4.62$ MeV and 
$\beta_3^P=0.388$~\cite{Spear89}.  Coupling to this state only shifted 
the barrier distribution in energy, leaving its shape largely unchanged, 
as expected for a state which has such a large excitation 
energy~\cite{Morton99,Hagino97b}.  In the other calculation, inclusion 
of the $2^+_1$ state in $^{34}$S and $\beta_2^T=0.338$ and 
$\beta_6^T=-0.025$ for $^{168}$Er, the 
theoretically predicted sign for the hexacontatetrapole deformation, 
produced a barrier distribution that in no way resembled the measured 
shape (calculation not shown).  No reasonable combination of coupling 
to states in the projectile and the predicted value for $\beta_6^T$ in
$^{168}$Er could be found to reproduce the desired shape of the measured 
barrier distribution.

\subsubsection{Coupling to transfer channels}

Other weaker couplings not yet taken into account include transfer 
channels.  The effective $Q$ values, that is after taking into account
the change in $Z_1 Z_2$ of the transfer products (where relevant), for 
various transfer channels are shown in Table~\ref{Table3}.  
Most of these transfer channels have a negative effective $Q$ value, 
and the effect of their inclusion in a CC calculation would be to 
produce barriers with energies greater than the average fusion 
barrier~\cite{Dasso83a}.
However, in the presence of the strong collective couplings in 
$^{168}$Er, the influence of the negative effective $Q$ value transfers 
on the shape of the barrier distribution is likely to be 
small~\cite{Leigh95}, as was seen with the inclusion of the inelastic
excitations in $^{34}$S in Sec.~\ref{proj}.  With the CC code used in 
this analysis it was not possible to calculate correctly the effects of 
all the above transfer channels.  However, a calculation was performed 
including the $2$n pickup 
channel, since the macroscopic form factor assumed for this pair transfer 
is likely to be close to the results from a microscopic 
treatment~\cite{Dasso85}.  
Coupling to the pair transfer was made using the form factor 
\begin{equation}
F_{\mbox{\scriptsize tran}}(R,\theta) = - \sigma_t (1+\beta_2^T Y_{20}
+ \beta_4^T Y_{40} + \beta_6^T Y_{60} ) \frac{d V (R,\theta)}{dR},
\label{tran} 
\end{equation}
where $R$ is the radius of the deformed target nucleus, $\theta$ is
the orientation angle of the deformed (axially symmetric) target 
nucleus, and $Y_{\lambda 0}$ are the spherical harmonics.  
In Eq.\ \ref{tran}, $\sigma_t$ is the transfer strength parameter.

\begin{figure}
\epsfxsize=18cm
\epsfbox[80 50 510 465]{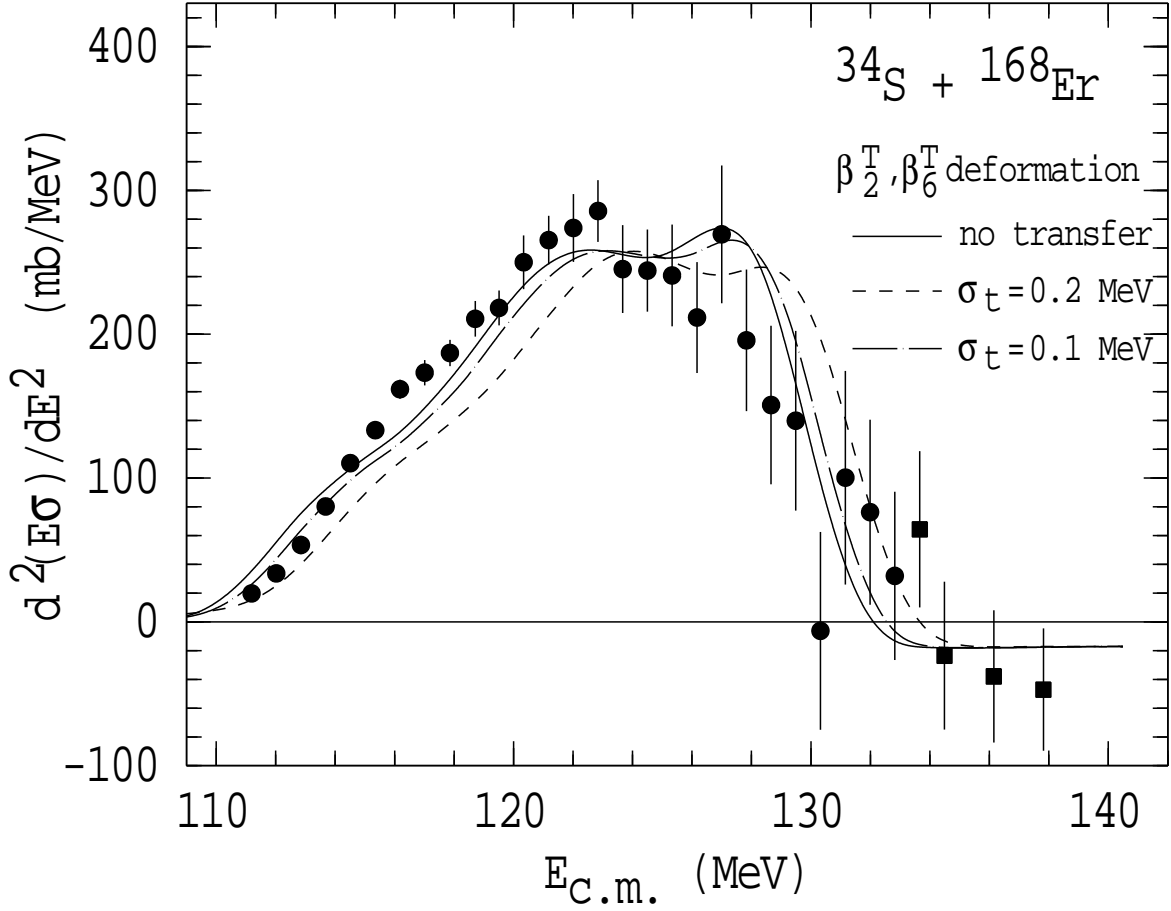}
\caption{Barrier distributions with \protect$\beta_2^T=0.338$ and 
\protect$\beta_6^T=+0.025$ and the \protect$2$n pickup transfer channel
with \protect$\sigma_t=0.2$ MeV (dashed line), \protect$\sigma_t=0.1$ 
MeV (dot-dashed line), and without the transfer channel (solid line). 
Note that none of the barrier distributions in this figure have been 
shifted in energy.  }
\label{fig7}
\end{figure}

The results are shown by the dashed line 
in Fig.\ \ref{fig7} which, in addition to the $\beta_2^T$ and 
$\beta_6^T$ deformation, with $\beta_2^T=0.338$ and $\beta_6^T=+0.025$, 
includes coupling 
to the 2n transfer channel with $Q=+2.7$ MeV and $\sigma_t=0.2$ MeV, 
the latter based on previous estimates of the pair transfer coupling 
strength~\cite{Broglia85,Dasso85}.  The effect of the positive $Q$ 
2n pickup channel is to redistribute some barrier strength in 
the barrier distribution, although the overall shape of the barrier 
distribution is not changed significantly.  
For comparison, the dot-dashed
line in Fig.\ \ref{fig7} shows the effect on the barrier distribution
when the transfer coupling strength is reduced to $\sigma_t=0.1$ MeV.
An increase in the transfer strength parameter resulted in an increase 
in the width of the barrier distribution, producing a distribution wider 
than the experimental one. 
A calculation with 2n pickup channel and the theoretically predicted 
sign for $\beta_6^T$ 
resulted in barrier distribution that was even more peaked than that 
shown by the dot-dashed line in Fig.\ \ref{fig4}(c).  Neither of these 
calculations are shown.  

As found in Sec.\ \ref{proj}, no reasonable combination of weak 
coupling, in this case the 2n transfer channel, and the theoretically 
predicted sign for $\beta_6^T$ (ie negative sign) could mimic the shape 
of the barrier distribution obtained for a positive value of $\beta_6^T$.
It seems unlikely that coupling to known additional transfer channels 
could change the shape of the $^{34}$S + $^{168}$Er barrier distribution 
to the extent the $\beta_6^T$ in $^{168}$Er does and thus account for 
the barrier distribution shape without resorting to the inclusion of 
$\beta_6^T$ in the CC calculations.

\section{Summary and Conclusion}
\label{conclusion}

Detailed coupled-channels calculations have been performed in an 
attempt to describe the shape of the recently measured barrier 
distribution~\cite{Morton00let,Morton00} for the $^{34}$S + $^{168}$Er 
reaction.  The $^{168}$Er
nucleus was chosen to investigate the disagreement between
theory and measurement at energies $1$ or $2$ MeV below the average
barrier observed in a range of reactions on deformed target nuclei.
Since $^{168}$Er has a very small hexadecapole 
deformation, and because the $3^-_1$ state in $^{168}$Er couples
very weakly, it was expected that the presence of the quadrupole 
deformation alone would allow for a stringent test of the fusion model. 

The calculated shape of the barrier distribution from a coupled 
channels calculation which included the quadrupole deformation only, 
failed to match the experimental $^{34}$S + $^{168}$Er barrier 
distribution.  
Agreement with the experiment was improved significantly when the 
hexacontatetrapole deformation was included, with $\beta_6^T=+0.025$.
However this value for $\beta_6^T$ has a sign opposite to that 
predicted in the macroscopic-microscopic model of M\"{o}ller 
{\em et al.}\ \cite{Moeller95}.

When the theoretically predicted value for $\beta_6^T$ was used, this 
resulted in a double-peaked barrier distribution in contrast with the 
data.  No combination of coupling to low-lying states in the projectile 
and/or the 2n transfer channel, when using the theoretically predicted 
value for $\beta_6^T$, could reproduce the shape of the experimental 
barrier distribution.  The best reproduction was obtained with a 
positive value for $\beta_6^T$.  It was argued that if 
such higher-order deformations were to be visible in the fusion barrier 
distribution, then the $^{168}$Er nucleus is likely to be one of the 
best candidates to observe their presence, since it has a very
small hexadecapole deformation.  

Within the framework of current fusion models, the fusion barrier 
distribution for $^{34}$S + $^{168}$Er apparently defines the $\beta_6^T$
deformation of $^{168}$Er as positive, in contrast with theoretical
predictions.  Further theoretical investigations are warranted, which
could also examine the possible influence on fusion of other higher-order 
deformations not considered in this analysis.

\begin{table}
\begin{center}
\caption{\small Parameters for the real nuclear potential for 
\protect$^{34}$S 
+ \protect$^{168}$Er 
[see \protect Eq.\ (1)].  }

\begin{tabular}{lcc}                      
$V_0$ (MeV)  &  $r_0$ (fm)    &  $a$ (fm)  \\ \hline
     $392.5$ &  $0.80$        &  $1.35$    \\
     $292.1$ &  $1.10$        &  $0.65$    \\ \hline
\end{tabular}
\label{table1}
\end{center}
\end{table}

\begin{table}
\begin{center}
\caption{\small Summary of the \protect$\beta_2^T$, $\beta_4^T$, and 
$\beta_6^T$ deformation parameters for nuclei as indicated. }
\begin{tabular}{lccc}
Nucleus     & $\beta_2^T$  & $\beta_4^T$  &  $\beta_6^T$   \\ \hline
$^{168}$Er  & $+0.338$~\cite{Raman87} &  $-0.007$~\cite{Moeller95} 
   &  $-0.025$~\cite{Moeller95}  \\
$^{154}$Sm  & $+0.28$~\cite{Leigh95}   &  $+0.05$~\cite{Leigh95}  
   &  $-0.005$~\cite{Moeller95}     \\
$^{238}$U   & $+0.275$   &  $+0.05$   &  $-0.015$~\cite{Moeller95} 
\\ \hline
\end{tabular}
\label{Table2}
\end{center}
\end{table}

\begin{table}
\begin{center}
\caption{\small Summary of the effective \protect$Q$ values 
\protect$Q_{\mbox{\scriptsize eff}}$, for various 
transfer channels for the reaction 
\protect$^{34}$S + \protect$^{168}$Er. } 
\begin{tabular}{lll}
Transfer products  & channel &  \protect$Q_{\mbox{\scriptsize eff}}$ 
\\ \hline
$^{35}$S + $^{167}$Er   &   n pickup      &        $-0.79$   \\
$^{33}$S + $^{169}$Er   &   n stripping   &        $-5.4$    \\
$^{35}$Cl + $^{167}$Ho   &   p pickup      &       $-7.4$    \\
$^{33}$P + $^{169}$Tm   &   p stripping   &        $+0.74$   \\
$^{36}$Cl + $^{166}$Ho   &   d pickup      &       $-6.2$    \\
$^{32}$P + $^{170}$Tm    & d stripping    &        $-2.8$    \\
$^{36}$S + $^{166}$Er    & 2n pickup      &        $+2.7$    \\
$^{32}$S + $^{170}$Er   &  2n stripping   &        $-6.8$    \\ \hline
\end{tabular}
\label{Table3}
\end{center}
\end{table}

\end{document}